\documentclass[conference]{IEEEtran}

\usepackage[pdftex]{graphicx}
\usepackage[cmex10]{amsmath}
\usepackage{array}
\usepackage{mdwmath}
\usepackage[tight,footnotesize]{subfigure}
\usepackage{url}
\usepackage{amssymb}
\usepackage{bbm}
\usepackage{amsfonts}
\usepackage{color}
\usepackage{cite}

\IEEEoverridecommandlockouts

\IEEEoverridecommandlockouts
\ifCLASSINFOpdf
\else
\fi
\begin{document}

\title{Performance Evaluation of Flow Allocation with Successive Interference Cancelation for Random Access WMNs}

\author{\IEEEauthorblockN{Nikolaos Pappas}
\IEEEauthorblockA{Department of Science and Technology\\
Link\"{o}ping University, Norrk\"{o}ping SE-60174,\\
Sweden\\
Email: nikolaos.pappas@liu.se}
\and
\IEEEauthorblockN{Manolis Ploumidis}
\IEEEauthorblockA{Institute of Computer Science\\
Foundation for Research and\\
Technology, Hellas (FORTH)\\
Email: ploumid@ics.forth.gr}
\and
\IEEEauthorblockN{Apostolos Traganitis}
\IEEEauthorblockA{Institute of Computer Science\\
Foundation for Research and\\
Technology, Hellas (FORTH)\\
Email: tragani@ics.forth.gr}}

\maketitle

\begin{abstract}
In this study we explore the performance gain that can be achieved at the network level
by employing successive interference cancelation (SIC) instead of treating interference as noise for
random access wireless mesh networks with multi-packet reception capabilities.
More precisely we explore a distributed flow allocation scheme aimed at maximizing average aggregate flow throughput
while also providing bounded delay combined with SIC.
Simulation results derived from three simple topologies show that the gain over treating interference as noise
for this scheme can be up to $15.2\%$ for an SINR threshold value equal to $0.5$.
For SINR threshold values as high as $2.0$ however, this gain is either insignificant or treating
interference as noise proves a better practice. 
The reason is that although SIC improves the throughput on a specific link, it also increases the interference imposed on neighbouring receivers.
We also show that the gain of applying SIC is more profound in cases of a large degree of asymmetry among interfering links.
\end{abstract}

\normalsize

\IEEEpeerreviewmaketitle

\section{Introduction}

In order to meet the increased demand for QoS over wireless mesh networks, a large number of studies has suggested aggregating network resources
by utilizing multiple paths in parallel. Multipath utilization however is an intriguing issue due to interference among neighbouring transmitters.

Different types of schemes have been suggested  that employ multiple paths in parallel including routing ones \cite{6133896}
or schemes that perform joint scheduling with routing, power control or channel assignment \cite{1717611,5501845,6130552}.
As far as flow allocation on multiple paths and rate control is concerned, a well studied approach associates a utility function to each flow's rate
and aims at maximizing the sum of these utilities subject to cross-layer constraints.
Several studies suggest joint rate control and scheduling approaches \cite{1430253, 4509706, conf_icc_QiuBX12}.
Authors in \cite{4712692} discuss a new fairness criterion and a new max-min fairness definition for multihop wireless networks.
Authors in \cite{5089987}, instead of employing a utility function of a flow's rate, they employ a utility function of flow's effective
rate in order to take into account the effect of lossy links.

As far as wireless random access networks are concerned, authors in \cite{1603389, Wang:2005:CRC:1062689.1062710} suggest cross layer designs that perform joint rate and MAC layer control
while \cite{7033238} explores the delay of a flow allocation scheme aimed at maximizing aggregate flow throughput.

Numerous studies have explored the performance in terms of several metrics, such as, delay, throughput, or delivery probability for schemes
that employ multiple paths in parallel along with some form of redundancy (i.e., network coding, diversity coding)
\cite{b:tsirig1, 6134700, 6335387, ref:papas_end_to_end, ref:papas_hop_by_hop}.

The performance in terms of throughput and delay in multi-user relay assisted wireless networks
studied in \cite{b:Pappas-arXiv-full-duplex} and \cite{b:Papadimitriou-arXiv}.

The severe effect of interference however on network performance is even more prominent when multiple paths are utilized in parallel.
Successive interference cancelation (SIC) is a promising physical layer technique for handling interference
and improving network performance \cite{4276942,Verdu:1998:MD:521411, 1210740, 5937210, 6339119}.
In \cite{5960833}, the performance of TDMA-based, conflict-free, scheduled multi-hop networks is studied when SIC is enabled at either all or at some nodes.
In \cite{b:TITHaenggi} a framework to study the performance of SIC in wireless networks using tools from Stochastic Geometry is provided.
A comprehensive survey on the performance of SIC for single- and multiple-antenna OFDM and spread OFDM (OFCDM) systems is provided in \cite{ b:SICSurvey}.
Authors in \cite{ b:TMC13} study the extent of throughput gains with SIC from a MAC layer perspective, and a SIC-aware scheduling algorithm is proposed.
The authors in \cite{b:PappasITW13} study the maximum stable throughput region for the two-user interference channel,
the case that the receivers perform SIC is also considered there.

In this study we explore the performance gain the can be achieved at the network level when SIC is employed
instead of treating interference as noise for random access wireless mesh networks with multi-packet reception
capabilities. More precisely we explore both the delay and  throughput
of a distributed flow allocation scheme suggested in our prior work \cite{6824997, DBLP:journals/corr/PloumidisPT14}
when combined with SIC. This scheme is aimed at maximizing average aggregate flow throughput
while also providing bounded delay. For the purposes of the evaluation process Ns2 simulation results from three simple topologies are derived.
Our results show that the flow allocation scheme discussed achieves up to 15.2\% higher AAT when combined with SIC instead of treating interference as noise
for an SINR threshold ($\gamma$) value equal to 0.5. For larger $\gamma$ values this improvement either becomes negligible or lower AAT
is achieved. This is due to the fact that the increased interference caused by links whose success probability is significantly
improved with SIC is not compensated by the gain in terms of throughput. Moreover the improvement in terms of throughput by employing SIC instead
of treating interference as noise increases with the asymmetry among interfering links.

\section{System Model}
\label{sec:system_model}

\subsection{Network Model}

We consider static wireless multi-hop networks with the following properties:
\begin{IEEEitemize}
    \item Random access to the shared medium where each node transmits independently of all other nodes based on its transmission probability only
    requiring no coordination among them.
    For flow originators transmission probability denotes the rate at which they inject packets into the network (flow rate).
    For the relay nodes transmission probability is fixed to a specific value and no control is assumed.
    \item Time is slotted and each packet transmission requires one time slot.
    \item Flows among different pairs of source and destination nodes carry unicast traffic of same-sized packets.
    \item All nodes are equipped with multi-user detectors thus can successfully decode packets
    from more than one transmitter at the same slot \cite{Verdu:1998:MD:521411}.
    \item We assume that all nodes are half-duplex and thus, cannot transmit and receive simultaneously.
    \item We also assume that all nodes always have packets available for transmission.
    \item As far as routing is concerned, multiple disjoint paths are assumed to be available by the routing protocol, one for each flow.
    Moreover, source routing is assumed ensuring that packets of the same flow are routed to the destination along the same path.
    Apart from that, for each node its position, transmission probability or flow rate along with an indication of whether it is a flow originator are assumed
    known to all other nodes. This information can be periodically propagated throughout the network through a link-state routing protocol.
\end{IEEEitemize}

\subsection{Channel Model}
\label{sec:chan_model}

The multi-packet reception (MPR) channel model used in this paper is a generalized form of the packet erasure model \cite{Pappas:2014:SPI:2611842.2612059}.

A block fading channel model is considered here with Rayleigh fading, i.e. the fading coefficients $h_{ji}$ remain constant during one timeslot, but
change independently from one timeslot to another based on a circularly symmetric complex Gaussian distribution with zero mean and unit variance.
The noise is assumed to be additive white Gaussian with zero mean and unit variance. With $p_j$ we denote the transmission power of node $j$, and $r_{ji}$ is the
distance between transmitter $j$ and receiver $i$ with $a$ being the path loss exponent.

Let $\mathcal{D}^\mathcal{T}_{j,i}$ denote the event that node $i$ is able to decode the packet transmitted from node $j$ given a set of active transmitters
denoted by $\mathcal{T}$.
For the topology presented in Fig. \ref{fig:wireless_scenario} for example,
$\mathcal{D}^{\{ 1,2 \} }_{1,R}$ denotes the event that the relay ($R$) can decode the information from the first node when
nodes $1$ and $2$ are active ($\mathcal{T} = \{1,2\}$). When only $j$ is active the event $\mathcal{D}^{\{ j \} }_{j,i}$ is defined as
\begin{equation}
\mathcal{D}^{\{ j \} }_{j,i} \triangleq \left\lbrace R_j \leq \log_2 \left(1+ |h_{ji}|^2 r^{-a}_{ji} p_j \right) \right\rbrace,
\end{equation}
which is equivalent to $\mathcal{D}^{\{ j \} }_{j,i} = \left\lbrace 2^{R_j} - 1 \leq |h_{ji}|^2 r^{-a}_{ji} p_j \right\rbrace$.

For convenience we define $ \mathrm{SNR}_{ji} \triangleq |h_{ji}|^2 r^{-a}_{ji} p_j$ and $\gamma_{j} \triangleq 2^{R_j} - 1$. The probability that the link $ji$ is not in outage when only $j$ is active is given by~\cite{b:Tse}
\begin{equation} \label{eq:SNR}
\mathrm{Pr}\left(\mathcal{D}^{\{j\}}_{j,i}\right) =\mathrm{Pr} \left\lbrace \mathrm{SNR}_{ji} \geq \gamma_{j} \right\rbrace = \exp \left(- \frac{\gamma_j r^{a}_{ii}}{p_{j}}\right).
\end{equation}

Let us consider the case that the relay node R treats interference from node 2 as noise when both nodes 1 and 2 are active. 
The event $\mathcal{D}^{\{ 1,2 \} }_{1R}$ is given by
\begin{equation}
\mathcal{D}^{\{ 1,2 \} }_{1,R} \triangleq \left\lbrace R_1 \leq \log_2 \left(1+ \frac{|h_{1R}|^2 r^{-a}_{1R} p_1}{1+|h_{2R}|^2 r^{-a}_{2R} p_2} \right) \right\rbrace ,
\end{equation}
which is equivalent to
\begin{equation}
\mathcal{D}^{\{ 1,2 \} }_{1,R} = \left\lbrace \gamma_1 \leq \frac{|h_{1R}|^2 r^{-a}_{1R} p_1}{1+|h_{2R}|^2 r^{-a}_{2R} p_2} \triangleq \mathrm{SINR}_1 \right\rbrace.
\end{equation}

The probability that the channel $1-R$ is not in outage when both nodes $1$ and $2$ are active is given by~\cite{b:Tse}:
\begin{equation} \label{eq:SINR_IAN}
\begin{aligned}
\mathrm{Pr}^{IAN}\left(\mathcal{D}^{\{ 1,2 \} }_{1,R}\right) =\mathrm{Pr} \left\lbrace \mathrm{SINR}_1 \geq \gamma_1 \right\rbrace = \\ = \exp \left(- \frac{\gamma_1 r^{a}_{1R}}{p_{1}} \right) \left[1+\gamma_1 \frac{p_{2}}{p_{1}} \left( \frac{r_{1R}}{r_{2R}} \right)^a \right]^{-1}.
\end{aligned}
\end{equation} 

Let us consider the case that the relay node R deploys successive interference cancelation (SIC) when both nodes $1$ and $2$ are active. 
If the relay R knows the codebook of the node $2$, it can perform SIC by first decoding the message sent by $2$, removing
its contribution (interference) to the received signal, and then decoding  the message coming from node $1$.
The relay $R$ is able to decode the interference, when both nodes $1$ and $2$ are active, if the following conditions are satisfied
\begin{align}
R_2 \leq \log_2 \left(1+ \frac{|h_{2R}|^2 r^{-a}_{2R} p_2}{1+|h_{1R}|^2 r^{-a}_{1R} p_1} \right), \\
R_1 \leq \log_2 \left(1+ |h_{1R}|^2 r^{-a}_{1R} p_1 \right),
\end{align}
which are equivalent to
\begin{align}
\gamma_2=2^{R_2} -1 \leq \frac{|h_{2R}|^2 r^{-a}_{2R} p_2}{1+|h_{1R}|^2 r^{-a}_{1R} p_1} \triangleq \mathrm{SINR}_{2R}\text{ and } \gamma_1 \leq \mathrm{SNR}_1.
\end{align}

The event $\mathcal{D}^{\{1,2\}}_{1,R}$ is given by $\mathcal{D}^{\{1,2\}}_{1,R} = \left\lbrace \mathrm{SINR}_{2R} \geq \gamma_2 \right\rbrace \cap \left\lbrace \mathrm{SNR}_1 \geq \gamma_1 \right\rbrace$,
and the probability that $R$ can decode the transmitted information from $1$ (given that both $1$ and $2$ are active) is given by (\ref{eq:SINR_SIC}) \cite{b:PappasITW13}.

{ \scriptsize
\begin{equation}\label{eq:SINR_SIC}
\begin{aligned} 
&\mathrm{Pr}^{SIC}\left(\mathcal{D}^{\{1,2\}}_{1,R}\right) =\mathrm{Pr}\left\lbrace \left\lbrace \mathrm{SINR}_{2R} \geq \gamma_2 \right\rbrace \cap \left\lbrace \mathrm{SNR}_1 \geq \gamma_1 \ \right\rbrace \right\rbrace \\
&= \exp \left(- \frac{\gamma_1 r^{a}_{1R}}{p_{1}} \right)  \exp \left[- \frac{\gamma_2(1+\gamma_1) r^{a}_{2R}}{p_{2}} \right]  \left[1+\gamma_2 \frac{p_{1}}{p_{2}} \left( \frac{r_{2R}}{r_{1R}} \right)^a \right]^{-1}.
\end{aligned}
\end{equation}
}

For the rest of the paper, for reasons of brevity the probability that node $i$ is able to decode
the packet transmitted from node $j$ given a set of active transmitters denoted by $\mathcal{T}$ ($\mathrm{Pr}(\mathcal{D}^{\{ \mathcal{T} \} }_{j,i})$)
will be denoted by $p_{j/\mathcal{T}}^{i}$.

\section{Analysis}
\label{sec:analysis}

The method for formulating aggregate throughput optimal flow rate allocation as an optimization problem for random topologies is presented in detail
in our prior work \cite{6824997, DBLP:journals/corr/PloumidisPT14}. However a modification incorporated to the corresponding optimization problem
that concerns transmission probabilities at the relay nodes needs to be discussed. For that reason, some notations along with the
final form of the corresponding optimization problem are presented.

\textit{V} denotes the set of the nodes and $|V|=N$.
We assume $m$ flows $f_{1}, f_{2},...,f_{m},$ that need to forward traffic to destination node $D$.
The analysis that follows can also be applied for the case where multiple flows have different destination nodes.
$R = \lbrace r_{1}, r_{2}, ...,r_{m} \rbrace$ represents the set of $m$ disjoint paths employed by these flows.
$|r_{i}|$ is used to denote the number of links in path $r_{i}$.
$I_{i,j}$ is the set of nodes that cause interference to packets sent from \textit{i} to \textit{j}.
Further on, $Src(r_{k})$ is used to denote the source node of the $k^{th}$ flow employing path $r_{k}$
while $r(i)$ returns the index of the path where node $i$ belongs.
$\bar{T}_{i,j}$ and $\bar{T}_{r_{k}}$ denote the average throughput measured in packets per slot achieved by link (i,j) and flow $f_{k}$
forwarded over path $r_{k}$ respectively.
Let also $I_{i,j}[n]$ denote the id of the n$^{th}$ interfering node for link (i,j).
For each node $i$, $q_{i}$ denotes its transmission probability given that there is a packet
available for transmission in its queue. As already discussed, for flow originators it indicates the rate at
which flow is injected on a path while for relay nodes it is assumed fixed to a specific value.
Finally $P_{ r_{k} } = \prod_{ (i,j) \in r_{k} } p_{i/i}^{j}$ is used to denote the \textit{end-to-end success probability} for path $r_{k}$.

Average throughput for a random link (i,j), $\bar{T}_{i,j}$, can be expressed through (\ref{eq:process_step_1}).
\vspace{-0.094in}
\begin{equation}
\label{eq:process_step_1}
\bar{T}_{i,j}  =  \sum_{ l=0 }^{ 2^{ L_{i,j} }-1 } P_{i,j,l}  q_{i,j}  \prod_{n=1}^{ L_{i,j} } q_{I_{i,j}[n]}^{b(l,n)}  (1 - q_{I_{i,j}[n]} )^{1-b(l,n)},
\end{equation}
where
\begin{equation}
\label{eq:relay_tx_prob}
\begin{aligned}
& q_{i,j}=\left\{\begin{matrix}
& q''_{i} \quad & j =  D\\
& q''_{i}(1-q''_{j}) \quad & j\neq D\\
\end{matrix}\right.,\\
\end{aligned}
\end{equation}
\begin{equation} \label{eq:indic_function}
\begin{aligned}
& q''_{i}=\left\{\begin{matrix}
& q_{i} \quad &  \quad ,i  \ne relay\\
& q_{i}\mathbbm{1}[q_{Src(r(i))}>0]  &\quad , j = relay\\
\end{matrix}\right.\\
\end{aligned}
\end{equation}
\begin{equation*}
\begin{aligned}
& P_{i,j,l}=p_{i/i\cup \{ I_{i,j}[n], \; \forall \; n: \; b(l,n) \neq0) \} }^{j}, \\
& b(l,n) = l \; \& \; 2^{n-1}, \; \text{\& is the logical bitwise AND operator.} \\
\end{aligned}
\end{equation*}

In (\ref{eq:indic_function}), $\mathbbm{1}[q_{Src(r(i))}>0]$ denotes an indicator function whose value becomes one if $q_{Src(r(i))}>0$ and zero otherwise.
The reason for employing this indicator function is discussed in the end of this section.
Recall that for flow originators $q_{i}$ denotes flow rate.
As also described in section \ref{sec:sim_setup}, transmission probability and position for every node can be periodically propagated to all other nodes
through routing protocol's topology control messages.
Position information is used to infer each link's success probability based on equation \ref{eq:SINR_IAN} or \ref{eq:SINR_SIC} depending on whether interference
is treated as noise (IAN) or successive interference cancelation (SIC) is employed at the receiver.
The average aggregate throughput achieved by all flows is expressed through $\bar{T}_{aggr} = \sum_{ k= 1 }^{m} \bar{T}_{r_{k} }$
where $\bar{T}_{r_{k}} = \underset{ (i,j) \in r_{k} }{ min }  \bar{T}_{i,j}$.

Aggregate throughput optimal flow rates that also provide bounded packet delay for a set of flows and a specific wireless topology can be
estimated by solving the following optimization problem:
\begin{equation*}
\begin{aligned}
& \underset{S'}{\text{Maximize}} \sum_{k=1}^{m}\left\{\begin{matrix}
\bar{T}_{Src(r_{k}),D}, & |r_{k}|=1 \quad \quad \quad \quad \quad (P2)\\
q'_{Src(r_{k})}, & |r_{k}|>1 \quad \quad \quad \quad \quad \quad \quad \quad
\end{matrix}\right.\\
& s.t.: \\
& \quad \quad (S1): \; 0 \leq q_{Src(r_{k})} \leq 1, \; k=1,...,m\\
& \quad \quad(S2): \; \bar{T}_{ Src(r_{k}),i} \leq \bar{T}_{j,l}, \\
& \quad \quad \quad \quad \quad \lbrace \forall i,j,k,l :  (Src(r_{k}),i), \; (j,l) \in r_{k}, |r_{k}| > 1 \\
& \quad \quad \quad \quad \quad k=1,...,m \rbrace \\
& \quad \quad(S3): \; 0 \leq  q'_{Src(r_{k}) } \leq 1, \; \lbrace \forall k :  |r_{k}|>1 \rbrace \\
& \quad \quad(S4): \; q'_{Src(r_{k}) } \leq \bar{T}_{i,j}, \; \lbrace \forall i,j,k :  |r_{k}|>1, \; (i,j) \in r_{k} \rbrace \\
\end{aligned} \\
\end{equation*}
where, $S'=\lbrace q_{Src (r_{k} )}, \;  k=1,...,m \rbrace \cup \lbrace q'_{src (r_{k} )} : \;  |r_{k}|>1 \rbrace$.

In the above optimization problem, constraint set S1 ensures that the maximum data rate for any flow does not exceed one
packet per slot while also allowing paths that are
not optimal to use, to remain unutilized.
Constraint S2 ensures that the flow injected on each path, that is the throughput of that path's first link, is limited by the flow that
can be serviced by any subsequent link of that path. In this way data packets are prevented from accumulating at the relay nodes
providing thus bounded packet delay.
For the rest of the paper this constraint will be referred to as \textit{bounded delay constraint}.
Moreover, the scheme that determines the flow to be assigned on each path based on the above optimization problem
will be referred to as \textit{Throughput Optimal Flow Rate Allocation (TOFRA)} scheme for the rest of the study.

Going back to the indicator function in (\ref{eq:indic_function}), the reason for employing it is the following: assume that the flow assigned on
a path is zero packets per slot. This means that relay nodes along this path will have no packets to transmit to their next hops.
However while enumerating all interfering nodes for expressing a specific link's average throughput through (\ref{eq:process_step_1})
relay nodes that belong to a path to which zero flow is assigned will be assumed to contribute with interference. This is due to the assumption
mentioned in the system model that all nodes always have packets available for transmission. Employing however the indicative function present in
(\ref{eq:indic_function}) a relay node $i$ that belongs to a path where zero flow is assigned ($q_{Src(r(i))}=0$) will not be considered
to contribute with interference.

\section{Simulation Setup}
\label{sec:sim_setup}

We evaluate the proposed aggregate Throughput Optimal Flow Rate Allocation scheme (TOFRA)
using network simulator NS-2, version 2.34 \cite{ref:ns2}, including support for multiple transmission rates \cite{ref:dei80211mr}.

\begin{table}[t]
\caption{Parameters used in the simulations}
\label{tab:param_simul}
\begin{center}
\begin{tabular}{ll}
\hline
Parameter & Value\\ \hline
Max Retransmit Threshold & 3\\
Contention Window & 5\\
Path Loss Exponent & 3.0\\
Packet size & 1500 bytes\\
Simulation duration & 20.000 slots\\
Transmission power & 0.1 W\\
Noise power & $7 \times 10^{-11} W$\\
\hline
\end{tabular}
\end{center}
\end{table}

Concerning medium access control, a slotted aloha-based MAC layer is implemented.
Transmission of data, routing protocol control and ARP packets is performed at the beginning of each slot without performing carrier sensing prior to transmitting.
Acknowledgements for data packets are sent immediately after successful packet reception while failed packets are re-transmitted.
Slot length, $T_{slot}$, is expressed through: $T_{slot} = T_{data} + T_{ack} + 2D_{prop}$ where $T_{data}$ and $T_{ack}$ denote
the transmission times for data packets and acknowledgements (ACKs) while $D_{prop}$ denotes the propagation delay.
It should be noted that all packets have the same size shown in table \ref{tab:param_simul}.
All network nodes, apart from sources of traffic, select a random number of slots before transmitting drawn
uniformly from $[0,CW]$. The contention window (CW) is fixed for the whole duration of the simulation and equal to $5$.

As far as physical layer is concerned the success probability for a link
is estimated as follows: in case where a flow allocation scheme variant is simulated assuming
that receivers perform successive interference cancelation (SIC) then the probability that a packet transmitted along link $(j,i)$
is successfully received given than nodes in $\mathcal{T}$ are also active is estimated based on (\ref{eq:SINR_SIC}).
If instead interference is treated as noise, success probability for this case is derived by employing (\ref{eq:SINR_IAN}).
Transmitters during each slot, that are considered to cause interference, are those transmitting data packets or routing protocol control packets.
All nodes use the same SINR threshold. Transmission power and noise is $0.1$ Watt and $7 \cdot 10^{-11}$ Watt respectively
while the path loss exponent is assumed equal to $3.0$.

As far as routing is concerned, static predefined routes to the destination are employed.
Hello and Topology Control (TC) messages are propagated throughout the network every one and five seconds respectively.
Each topology control message may carry the following information: a)transmission probability b)position, and
c)an indication of whether it is a flow originator or not.
As also discussed in Section~\ref{sec:system_model}, transmission probabilities are assumed to be fixed for relay nodes since contention
window (CW) remains fixed for the whole simulation period.
Using this information from the TC messages each node can infer both the network topology and the success probability for each link
based on either (\ref{eq:SINR_IAN}) or (\ref{eq:SINR_SIC}) since all link distances are known.
Upon each TC message reception, each flow source is able solve the topology-specific instance of the flow allocation optimization problem
presented in Section \ref{sec:analysis}
using the simulated annealing method. In this way the flow rates (packets per slot) that should be assigned on each path
in order to achieve maximum average aggregate throughput are estimated along with the average aggregate throughput for all flows.
According to this process, flow rates are estimated on a distributed manner for all flow originators.

In each simulation scenario flows carrying constant bit rate UDP traffic are generated while simulation period is 20.000 slots.
Queues for flow originators are kept backlogged for the whole simulation period.

\section{Numerical and Simulation Results}
\label{sec:results}

The evaluation process presented in this section consists of three parts. In the first one we briefly discuss the accuracy of the model deployed
by the TOFRA scheme on capturing the average aggregate flow throughput (AAT).
In \cite{DBLP:journals/corr/PloumidisPT14} we also show that this model accurately captures the AAT observed in
simulated scenarios based on larger random scenarios.
In the second part we explore the gain in terms of throughput that can be achieved at the network level by combining TOFRA
flow allocation scheme with SIC.
Finally in the third part we discuss the effect of SIC on end-to-end flow delay.

\begin{figure}[t]
\centering
\includegraphics[scale=0.2]{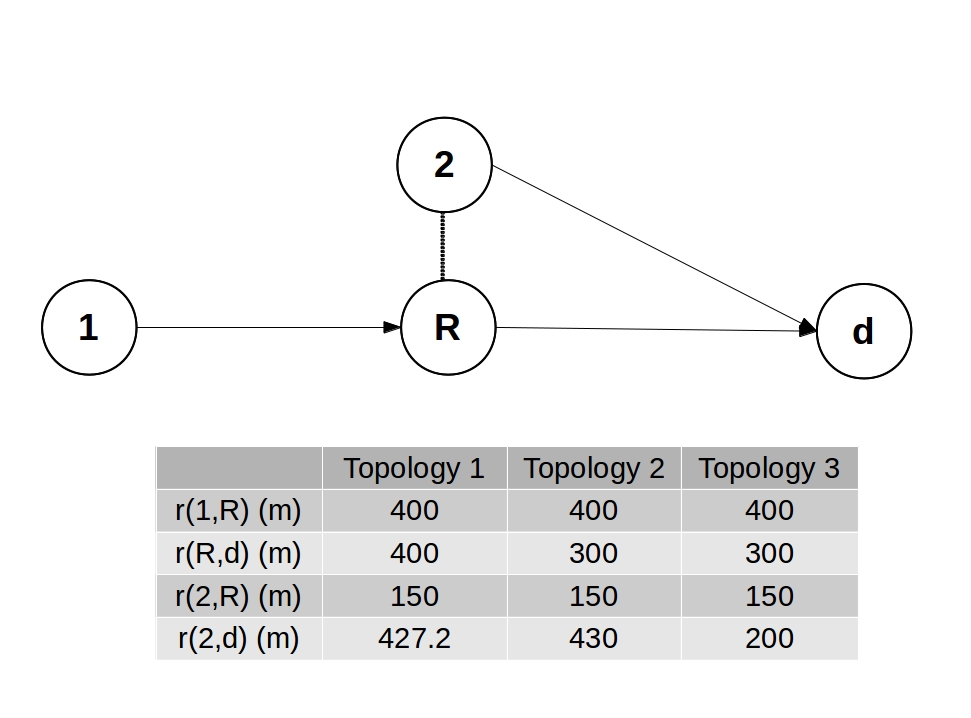}
\caption{Considered topologies with various link distances.}
\label{fig:wireless_scenario}
\end{figure}

For the rest of the paper the notion of asymmetry for two interfering links will be used to denote the difference between
the average received SNR/SINR over them.
As far as SIC is concerned it has been shown that performance gain increases with
the asymmetry among interfering links \cite{b:TMC13}. For that reason three different topologies are explored based on the one
presented in Fig. \ref{fig:wireless_scenario}. Different topology instances are derived by fixing the distances between pairs
of nodes. The corresponding distance values are summarized in the table incorporated in the same figure.
For all three topologies, two unicast flows are assumed sourced at nodes 1 and 2 respectively. Flow $f_{1}$ is forwarded to $d$
through path 1-R-d while flow $f_{2}$ through 2-d.
Assuming such a traffic scenario, in topology 1 depicted in Fig. \ref{fig:wireless_scenario}, transmissions along link (1,R) experience
interference from node $2$. If similar SINR threshold ($\gamma$) values for all transmitters are further assumed, then the received signal
on $R$ from $2$ constituting the interference is received with higher power compared to the signal received from $1$.
On a similar manner, in topologies 1 and 2 transmissions along link (2,d)
experience interference from $R$. The signal constituting interference from $R$ is received with higher power at $d$ than the
signal carrying data packets sent from $2$ (due to the different link distances).

Based on these remarks, different approaches are explored for each topology presented in Fig. \ref{fig:wireless_scenario} depending on how interference is handled
at each receiving node. For topologies 1 and 2 three different approaches are explored. In the first one interference at nodes $R$, $d$ is treated as noise.
In the second approach SIC is applied on $R$ as described in section \ref{sec:chan_model}.
In the third one destination $d$ first tries to decode the message from $R$, remove its contribution (interference) to the received signal
and then decode the message from $2$.
Finally as far as topology 3 depicted in Fig. \ref{fig:wireless_scenario} is concerned three approaches are also explored.
The first two approaches are the same with topologies 1 and 2. In the third one however where the destination resides closer to transmitting
node $2$ instead of $R$ so $d$ first tries to decode the message received from $2$ (interference) remove its contribution to the received signal
and then decode the message from node $R$.
For the rest of the paper we will also use the term successive interference cancelation to describe how interference is handled at destination $d$.
To distinguish among the different approaches discussed above for handling interference
they are labelled after: \textit{IAN}, \textit{SIC(R)}, \textit{SIC(R,d)} with SIC(R,d) denoting that SIC is applied at both $R$ and $d$.

As far as allocation of flow (data rates) on different paths is concerned three different schemes are explored. The first scheme is TOFRA
(Throughput Optimal Flow rate Allocation), proposed in our prior work \cite{6824997, DBLP:journals/corr/PloumidisPT14}
and presented in section \ref{sec:analysis}.
\textit{Full MultiPath (FMP)} assigns one packet per slot on each path. Finally the third scheme explored employs only a single path to forward traffic
to the destination. Based on how \textit{best} path is identified we explore two variants: in the first one denoted as $BP_{e2e}$, best path is
considered the as one exhibiting the highest end-to-end success probability (defined in section \ref{sec:analysis}).
In the second variant denoted as $BP_{wb}$, best path is defined as the one that has the \textit{widest} bottleneck link
which can be formulated as identifying path
$r_{k}$ : $\underset{k}{arg \; max} \; \underset{ (i,j) \in r_{k})}{Min} \; p_{i/i}^{j}$.
In the first two topologies explored $BP_{wb}$ utilizes path 1-R-d to the destination while in the third one 2-d.
$BP_{e2e}$ on the other hand deploys path 2-d for all three topologies explored.
Applying SIC for the topologies presented in Fig. \ref{fig:wireless_scenario} is meaningless since when path
2-d is used, destination $d$ receives no interference while in the case of 1-R-d the interference received at $d$ from $1$
is insignificant due to the large distance between them.
For both aforementioned best-path variants, the flow assigned on the utilized single path is calculated by solving a single-path version of the optimization problem (P2) presented in section
\ref{sec:analysis} using the simulated annealing method. 

For the purposes of the evaluation process different simulation scenarios are generated as follows: for each topology presented
in Fig. \ref{fig:wireless_scenario} one of the aforementioned flow allocation schemes is employed. For each flow allocation scheme three
variants are simulated based on how interference is handled at each receiving node. The variant denoted by \textit{FMP-IAN} for example
assigns one packet per slot on each path while interference is treated as noise at each receiver. For \textit{FMP-SIC(R)} SIC is assumed at receiving node R.

In the first part of the evaluation process we explore whether the model employed by the TOFRA scheme discussed accurately captures
the average aggregate flow throughput (AAT).
Table \ref{tab:tofra_num_sim} summarizes the flow rates assigned on each path along with the corresponding value for AAT achieved by TOFRA
derived from both the numerical and the simulation results.
Recall that flow rates assigned on each path are identified by sources by solving the topology specific instance of the
flow allocation optimization problem presented in section \ref{sec:analysis}.

\begin{table}[t]
\begin{center}
\begin{tabular}{|c|c|c|c|c|c|c|}
\hline
Topo & $\gamma$ & Flow alloc & $q_{1}$ & $q_{2}$ & $AAT_{num}$ & $AAT_{sim}$\\
 &  & scheme &  &  &  Pkts/Slot& Pkts/Slot\\ \hline
1 & 0.5 & TOFRA-IAN &  0.0 &  1.0 &  0.973 & 0.970 \\ \hline
1 & 0.5 & TOFRA-SIC(R) &  0.287 & 1.0  & 1.045 & 1.045 \\ \hline
1 & 0.5 & TOFRA-SIC(R,d) &  0.287 & 1.0 & 1.057 & 1.069 \\ \hline
1 & 2.0 & TOFRA-IAN & 0.0 & 1.0 & 0.896 & 0.891 \\ \hline
1 & 2.0 & TOFRA-SIC(R) &  0.164 & 1.0 & 0.783 & 0.802 \\ \hline
1 & 2.0 & TOFRA-SIC(R,d) & 0.164 & 1.0 & 0.782 & 0.824 \\ \hline \hline
2 & 0.5 & TOFRA-IAN & 0.0 & 1.0 & 0.972 & 0.968 \\ \hline
2 & 0.5 & TOFRA-SIC(R) & 0.350 & 1.0 & 1.005 & 1.004 \\ \hline
2 & 0.5 & TOFRA-SIC(R,d) & 0.350 & 1.0 & 1.084 & 1.116 \\ \hline
2 & 2.0 & TOFRA-IAN & 0.0 & 1.0 & 0.894 & 0.894 \\ \hline
2 & 2.0 & TOFRA-SIC(R) & 0.267 & 1.0 &  0.760 & 0.764 \\ \hline
2 & 2.0 & TOFRA-SIC(R,d) & 0.268 & 1.0 & 0.833  & 0.901 \\ \hline \hline
3 & 0.5 & TOFRA-IAN &  1.0 & 1.0 & 1.011 & 1.015 \\ \hline
3 & 0.5 & TOFRA-SIC(R) & 0.153 & 1.0 & 1.062 & 1.047 \\ \hline
3 & 0.5 & TOFRA-SIC(R,d) & 0.297 & 1.0 & 1.158 & 1.149 \\ \hline
3 & 2.0 & TOFRA-IAN &  0.0 & 1.0 & 0.988  & 0.985 \\ \hline
3 & 2.0 & TOFRA-SIC(R) & 0.060 & 1.0 & 0.915 & 0.954 \\ \hline
3 & 2.0 & TOFRA-SIC(R,d) & 0.232 & 1.0 & 1.006 & 1.001 \\ \hline
\end{tabular}
\end{center}
\caption{AAT (Pkts/Slot): Simulation vs Numerical results for each TOFRA variant}
\label{tab:tofra_num_sim}
\end{table}

The average deviation between the AAT derived from the model described in section \ref{sec:analysis} and
the one observed in the simulated results is 1.56\% over all topologies, $\gamma$ values and TOFRA variants employed.
There are several reasons for this deviations.
The main one is related to the assumption of the model for AAT concerning saturated queues at the relay nodes.
In our analysis it is assumed that whenever a relay node attempts to transmit a packet there is always one available at its queue.
In the simulated scenarios however, a relay node's queue may be empty at a specific slot.
In this way however the considered model for the AAT overestimates the interference experienced by any link in the simulated scenarios
and thus underestimates the average throughput achieved over that link. Due to the assumption concerning saturated queues at the relay nodes
it also overestimates the collision probability at each relay node due to concurrent packet transmission and reception events. 
At the end of this section we also discuss how this underestimation of a link's average throughput may also affect queueing delay.
Apart from that, in the analysis presented in section \ref{sec:analysis}, a packet is not assumed to be dropped afer a larger number of
failed retransmissions. In the simulation parameters presented in table \ref{tab:param_simul} however, a maximum retransmit threshold equal to $3.0$
is adopted. This suggests that after three failed transmissions a specific packet is dropped. This may result in lower throughput for the
link over which that packet is retransmit but will also result in reduced interference imposed on neighbouring links. Finally,
in the analysis we have assumed that whenever a packet is transmitted it is a packet carrying data.
In the simulated scenarios however, all nodes either perform periodic emission of routing protocol's control messages or forwarded specific
received control packets (topology control messages for the simulation setup presented in section \ref{sec:sim_setup}).
This means that specific slots are spent carrying routing protocol's control messages instead of data packets resulting
in our analysis overestimating the AAT observed in the simulated results.

\begin{figure}[t]
\centering
\subfigure[Topology 1]{
\includegraphics[width=8.8cm, height=6.5cm]{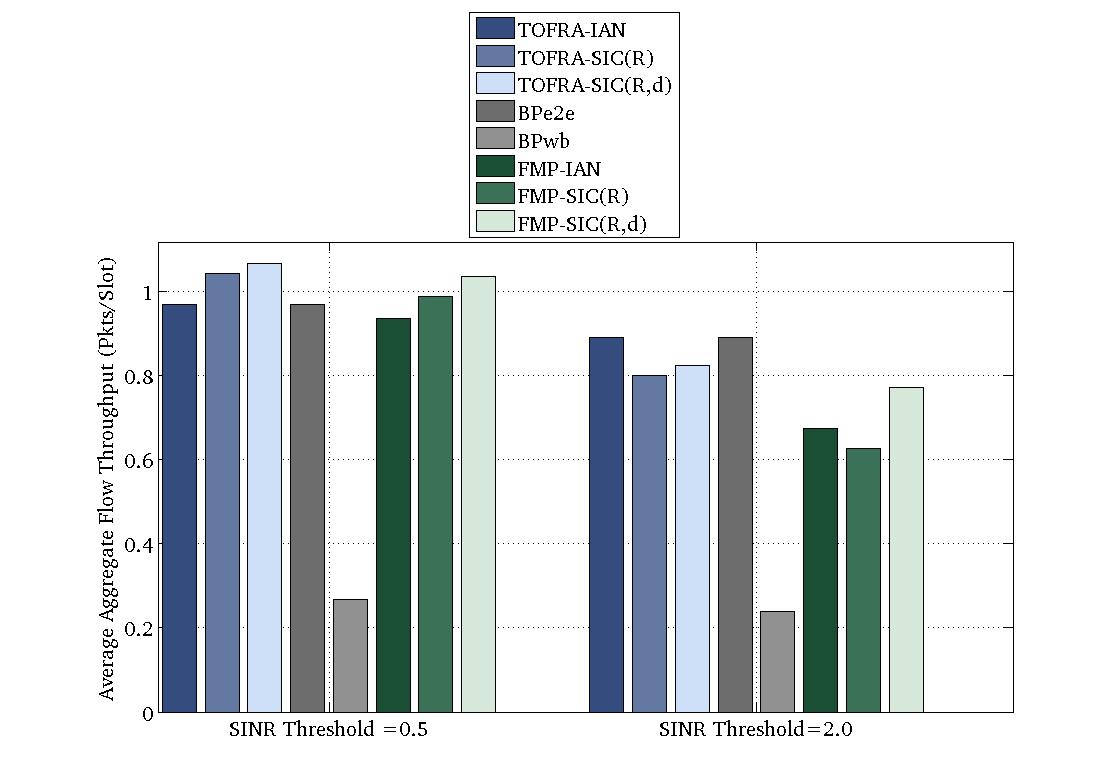}
\label{fig:aat_1}
}
\subfigure[Topology 2]{
\includegraphics[width=8.8cm, height=5.5cm]{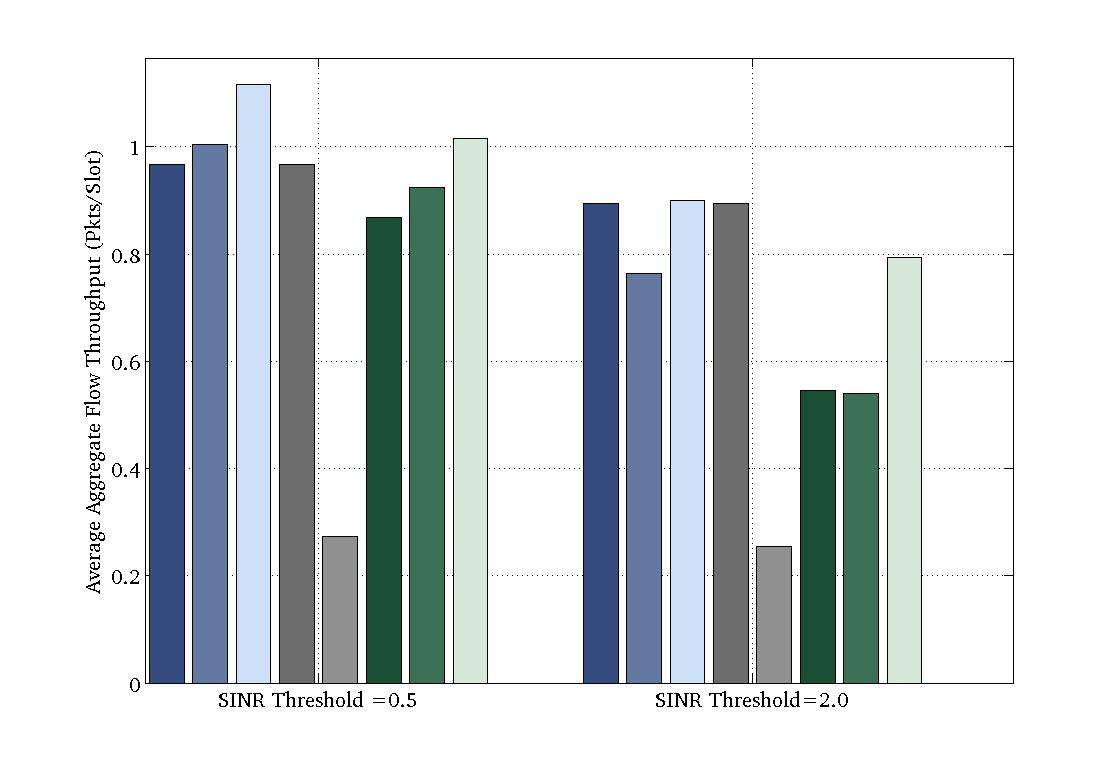}
\label{fig:aat_2}
}
\subfigure[Topology 3]{
\includegraphics[width=8.8cm, height=5.5cm]{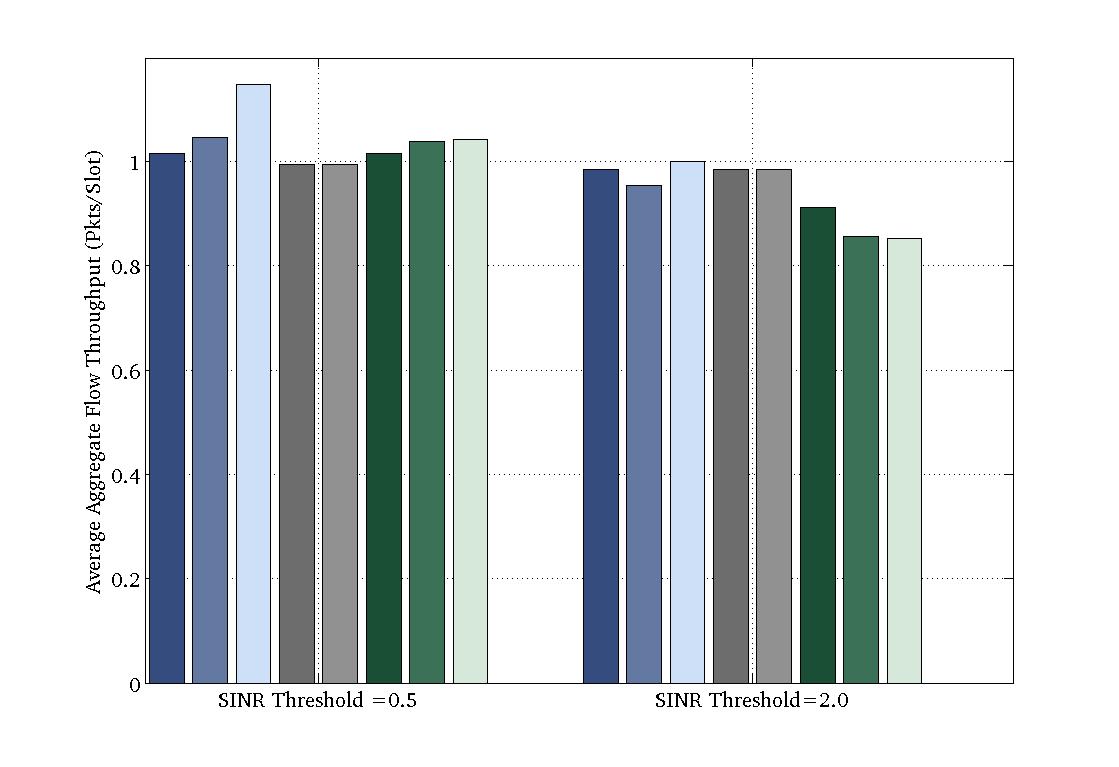}
\label{fig:aat_3}
}
\label{fig:flow_alloc_thrput}
\caption{Average Aggregate Throughput per Flow Alloc Variant}
\end{figure}

In the second part of the evaluation process we explore the gain in terms of throughput that can be achieved by deploying
SIC instead of treating interference as noise (IAN).
More precisely we explore the AAT achieved by the aforementioned flow allocation schemes when different approaches
for handling interference are followed (discussed above). Figs. \ref{fig:aat_1} - \ref{fig:aat_3} present the corresponding AAT values for the three topologies
summarized in Fig. \ref{fig:wireless_scenario}.

Figs. \ref{fig:aat_1}-\ref{fig:aat_3} show that applying SIC instead of IAN
at both receiving nodes $R$, $d$ proves gainful in terms of AAT when $\gamma$=0.5. For the case of the TOFRA flow allocation scheme the gain is
$10.2\%$, $15.2\%$, and $13.2\%$ respectively for the three topologies explored. The corresponding values for FMP are $10.7\%$, $16.9\%$,
and $2.6\%$ respectively. It should also be noted that the gain in terms of throughput for SIC is less significant when it is applied only
to receiver $R$. For $\gamma$=0.5 for example, employing SIC at $R$ instead of IAN results in $7.7\%$, $3.7\%$, and $3.1\%$ higher AAT for the three topologies explored.
Applying SIC on $R$ increases the success probability on link (1,R) from $9.3\%$ to $95.1\%$ for $\gamma$=0.5 and from $2.3\%$ to $81.5\%
$ for $\gamma$=2.0.
Consequently transmitter $1$ will manage to deliver a larger portion of its traffic to $R$ when SIC is employed at $R$ instead of IAN which will also
result in an increased number of packets transmitted from $R$ to $d$. This will have a negative effect on the average throughput of link (2,d) since
it will experience increased interference. Indeed, for the first topology presented in Fig. \ref{fig:wireless_scenario}, $\gamma$=0.5 and the TOFRA flow
allocation scheme, when interference is treated as noise at all the receivers the fraction of data packets transmitted over (2,d)
that are re-transmitted due to low SINR
is $2.7\%$. In the scenario where SIC is employed at $R$ the corresponding fraction of re-transmitted packets is $14.1\%$.
This shows that improving the success probability at a relay node by applying SIC will also increase the interference imposed on its next hop.
Consequently the number of failed packets that are re-transmitted will increase limiting the gain in terms of AAT.
As Figs. \ref{fig:aat_1}-\ref{fig:aat_3} also show, for $\gamma$ values as high as 2.0 applying SIC instead of IAN for the case of TOFRA
either offers an insignificant gain or results in lower average aggregate throughput. As already discussed, applying SIC at R significantly
increases the success probability on link (1,R) with TOFRA also increasing the amount of flow assigned on path 1-R-d.
If however the increased interference on link (2,d) is not compensated by the gain of utilizing path 1-R-d the average aggregate flow throughput (AAT)
observed may be lower compared to the case where IAN is applied at each receiver.

As far as the relation between interfering links asymmetry and gain in terms of throughput of SIC over IAN is concerned the following remark is also
interesting. As already discussed above, the success probability of link (1,R) increases from $9.3\%$ to $95.1\%$ for $\gamma$=0.5
when SIC is employed at $R$ instead of IAN. Accordingly in the topology 1 for example, the success probability
of link (2,d) increases from $60.4\%$ to $66.7\%$ for $\gamma$=0.5 when SIC is employed at $d$ instead of IAN. This increase in the success probability
is significantly lower than the corresponding one for link (1,R). The reason for this is the different asymmetry between interfering links for the two receivers.
As Fig. \ref{fig:wireless_scenario} also shows, the distance of interfering node $2$ from $R$ is much smaller than
the distance between $1$ and $R$. The distances however of nodes $2$ and $R$ from $d$ are very similar.
A notable effect of combining SIC with the TOFRA flow allocation scheme is the utilization of paths which where assigned zero flow when IAN was
applied at receiving nodes. As table \ref{tab:tofra_num_sim} shows, the utilization of path 1-R-d becomes non-zero for all topologies and $\gamma$ 
values considered when SIC is employed.

As far as different flow allocation schemes are concerned, variants of the TOFRA scheme achieve  higher AAT than the
corresponding full multi-path (FMP) variants for all topologies and $\gamma$ values employed. The main reason
for this is that FMP assigns one packet per slot on each path on an interference unaware manner resulting in a higher fraction of data packets
re-transmitted due to low SINR. This performance gap becomes even more profound when large SINR threshold values are assumed: thus,
the success probability of all links is decreased. Considering topology 3 with $\gamma$=0.5 for example, TOFRA-SIC(R,d) achieves $10.2\%$ higher
AAT than the corresponding FMP variant (FMP-SIC(R,d)).
Compared to $BP_{e2e}$, TOFRA-IAN achieves the same AAT for almost all scenarios explored since they both utilize at full rate path 2-d.
The only exception to this is the scenario based on topology 3 where $\gamma$=0.5. In this scenario the corresponding TOFRA variant also utilizes
path 1-R-d.
When TOFRA however is combined with SIC at both $R$
and $d$ and a low $\gamma$ is employed it achieves higher AAT. In topology 3 for example, TOFRA-SIC(R,d) achieves $15.4\%$ higher AAT than $BP_{e2e}$.
It should be noted however that the prospect of higher throughput for TOFRA is limited by the number of paths available. In \cite{DBLP:journals/corr/PloumidisPT14}
where several paths are employed in parallel we show that TOFRA outperforms $BP_{e2e}$.
Finally, best path variant that selects the path with the widest bottleneck link in terms of success probability ($BP_{wb}$) utilizes path 1-R-d for topologies 1 and 2
and 2-d for topology 3. As table \ref{tab:tofra_num_sim} shows however TOFRA assigns zero flow on path 1-R-d for most scenarios explored when interference is treated as noise.
This shows that utilizing this path in parallel with 2-d would result in lower AAT.
As also shown in Figs. \ref{fig:aat_1}-\ref{fig:aat_3}, $BP_{wb}$ achieves the lowest AAT among all schemes for topologies 1 and 2 and performs the same with
$BP_{e2e}$ for topology 3.

\begin{figure}[!t]
\centering
\subfigure[Topology 1]{
\includegraphics[scale=0.25]{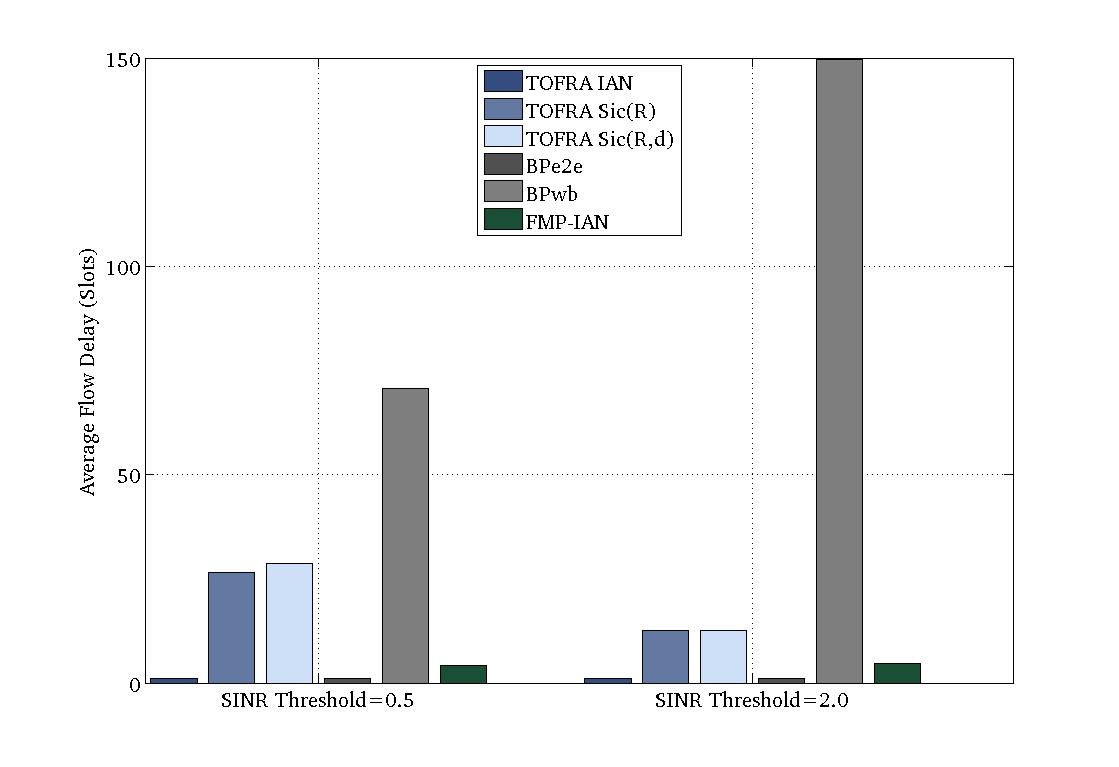}
\label{fig:delay_1}
}
\subfigure[Topology 2]{
\includegraphics[scale=0.25]{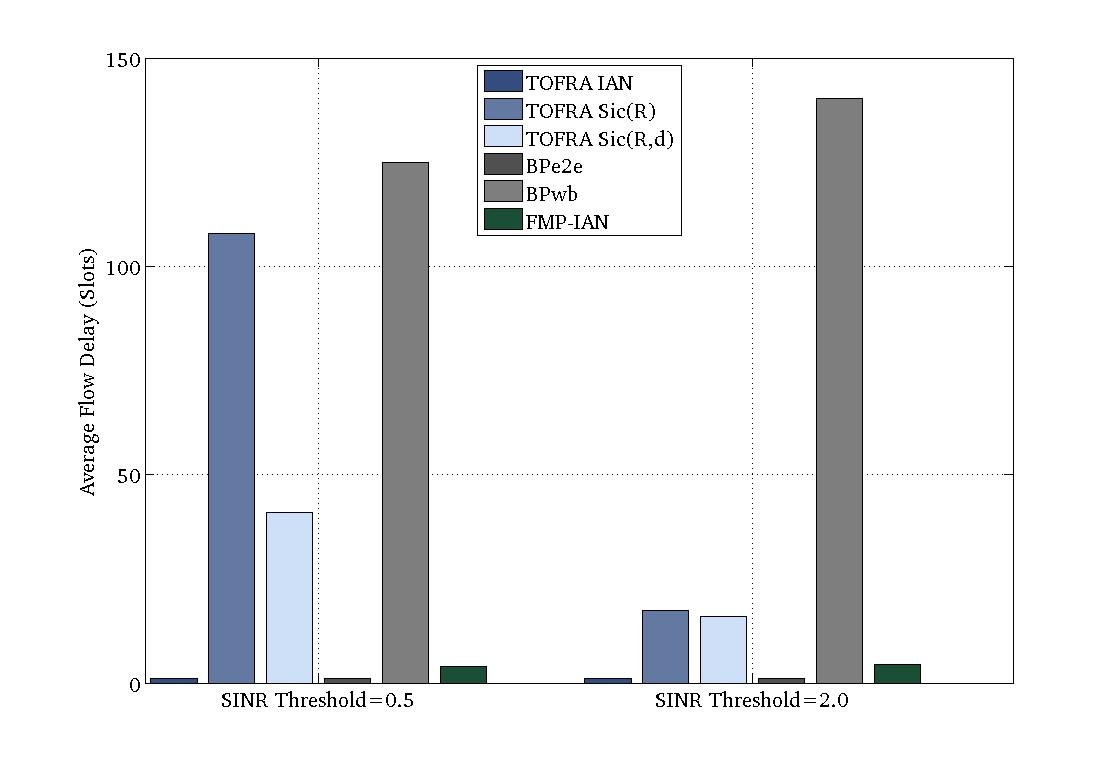}
\label{fig:delay_2}
}
\subfigure[Topology 3]{
\includegraphics[scale=0.25]{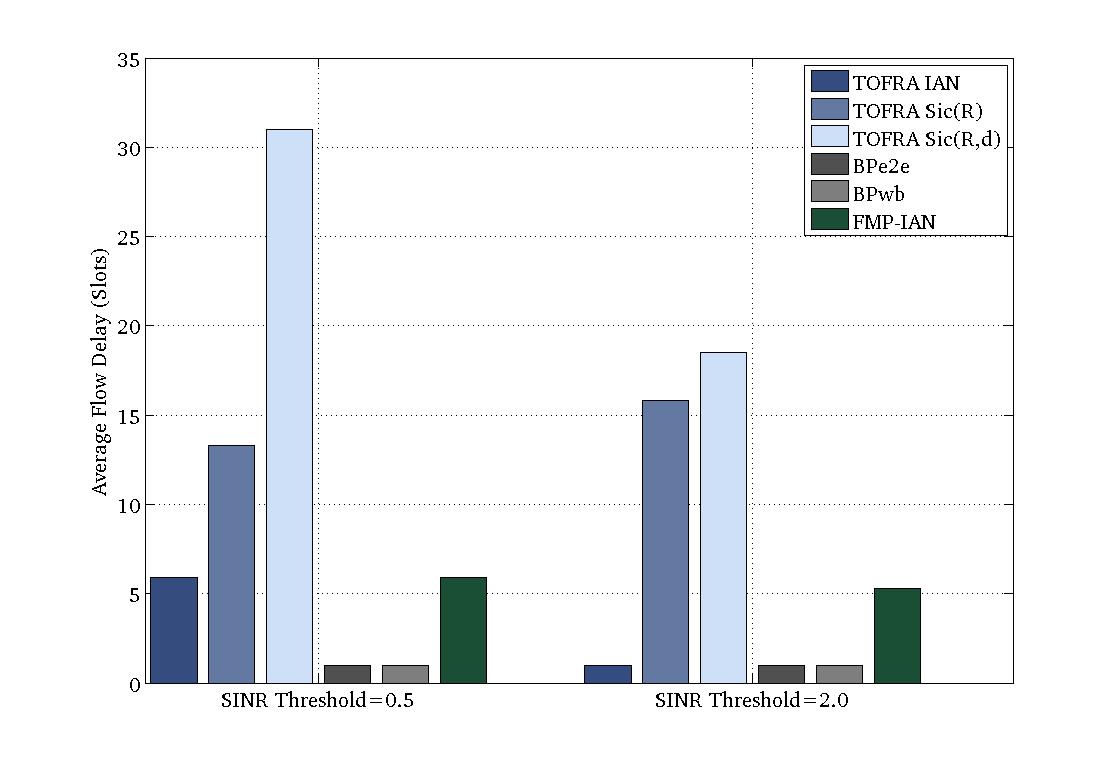}
\label{fig:delay_3}
}
\label{fig:delay}
\caption{Delay (Slots) per Flow Allocation Variant. }
\end{figure}

In the third part of the evaluation process the aforementioned flow allocation schemes are compared in terms of delay.
Moreover the effect of applying SIC on it is also explored. Before discussing simulation results the following definitions are necessary:
for each flow \textit{end-to-end flow delay} will be used to denote the average per packet end-to-end delay for all packets forwarded by that flow.
End-to-end delay for a packet is the interval between the time when it is transmitted for the first time at the source of the flow and the time
when the packet is successfully received at the destination of the corresponding flow.
Figs. \ref{fig:delay_1}-\ref{fig:delay_3} present average flow delay for the three topologies described in Fig.
\ref{fig:wireless_scenario} and SINR threshold values 0.5 and 2.0. For the rest of the paper end-to-end flow
delay will be referred to as \textit{flow delay}. Note that flow delay values for FMP are included only for the variant
where interference is treated as noise (FMP-IAN). When FMP is combined with SIC flow delay values are by far larger
than the corresponding values of the other schemes and including them in the aforementioned figures would constitute them illegible.
Instead flow delay values for FMP-SIC(R) and FMP-SIC(R,d) are discussed at the end of the section.

As these figures illustrate, for all three topologies explored, TOFRA-IAN achieves the same delay with $BP_{e2e}$ for both $\gamma$=0.5 and 2.0.
The only exception to this is the simulated scenario based on topology 3 with $\gamma$=0.5. In this scenario TOFRA-IAN also assigns flow on path 1-R-d.
The flow assigned on path 1-R-d is also increased in the case where TOFRA is combined with SIC. As table \ref{tab:tofra_num_sim}
for example shows, for the second topology and $\gamma$=0.5 TOFRA assigns 0.350 packets per slot on path 1-R-d when SIC is applied at both $R$ and $d$.
Consequently a larger number of packets will experience queueing delay at $R$ and also increased re-transmissions due inter-path interference.
The effect of queueing delay on flow delay is extensively discussed in the rest of the section.
This is also validated if $BP_{wb}$ is considered for topologies 1 and 2.
In these topologies, $BP_{wb}$ forwards all packets through path 1-R-d. Figs. \ref{fig:delay_1}-\ref{fig:delay_3} show that it experiences
significantly higher flow delay than all other schemes for these two topologies apart from full multipath variants FMP-SIC(R), FIM-SIC(R,d).
The second reason behind the increased delay of TOFRA when combined with SIC is related to the accuracy with which the model presented
in section \ref{sec:analysis} captures the average throughput of a link and is discussed in the next paragraph.

To validate the effect of queueing delay on flow delay the following quantities are also estimated: the \textit{throughput ratio}
for a relay node $R$ defined as $\bar{T}(1,R)/\bar{T}(R,d)$ where $\bar{T}(1,R)$ and $\bar{T}(R,d)$ denote the average
throughput for links (1,R) and (R,d) respectively. A value for that ratio larger than one would suggest a queue at the relay
where packets arrive at a rate faster than the rate they can be serviced (delivered to $d$). This results in an unstable queue at the relay node
and consequently on packets experiencing unbounded delay. A value for that ratio that is one would imply a sub-stable queue at $R$. 
In this case packets may experience increased queueing delay.
Additionally, the average queue length for each node, especially for relays, is calculated from simulated results.

Figs. \ref{fig:delay_1}-\ref{fig:delay_3} also show that for all topologies and $\gamma$ values explored FMP-IAN
achieves significantly lower flow delay than TOFRA
variants that employ SIC although FMP is expected to experience more failed packets due to increased inter-path interference.
In order to explore this delay gap between FMP and TOFRA variants, topology 2 with $\gamma$=0.5 is used as an example.
Moreover we focus on FMP-IAN and TOFRA-SIC(R,d) variants. Simulation results reveal that FMP-IAN indeed experiences a larger
percentage of failed transmissions due to low received SINR. For link (1,R) this ratio is $81.0\%$ for FMP-IAN and $3.62\%$ for TOFRA-SIC(R,d).
As far as path 1-R-d is concerned it is also interesting to note that FMP-IAN manages to deliver to R only $8.3\%$ of
the packets sent over link (1,R) while the corresponding value for TOFRA-SIC(R,d) is $68.6\%$.
Taking into account these ratios TOFRA-SIC(R,d) is expected to experience higher queueing delay than FMP-IAN.
Indeed, the average queue length for relay node $1$ is $0.09$ packets for the case of FMP-IAN and $17.0$ packets when TOFRA-SIC(R,d)
is employed.

Comparing different TOFRA variants in terms of average flow delay shows than when SIC is applied instead of IAN average flow delay
exhibits a significant increase. As table \ref{tab:tofra_num_sim} shows, when TOFRA is combined with SIC, maximum AAT is achieved
by utilizing path 1-R-d in parallel with 2-d for all topologies explored.
The gap in terms of delay may imply that TOFRA variants that employ SIC experience increased queueing delay.
To validate this the simulation scenario based on topology 2 with $\gamma$=0.5 is used as an example.
First the throughput ratio for relay node $R$ is estimated for TOFRA variants that employ SIC from simulation results.
For both variants the value of this ratio is $1.02$, and $1.01$ respectively  suggesting that the queue at $R$ becomes unstable.
However as already discussed in the first part of the evaluation process the model
employed for the average aggregate flow throughput may underestimate the actual average throughput of a link observed in the simulation scenarios.
In this way the average throughput of a specific link may by slightly higher than the average throughput of a subsequent link which
results in an \textit{unstable} queue at the relay. Part of our future work however is to relax the bounded delay constraint so as
to account for the deviation in capturing the actual average throughput of a link.
The second reason is that SIC improves the success probability of link (1,R) and thus the number of packets that are successfully
delivered to $R$ when compared to TOFRA-IAN resulting in a larger average queue size.

For all topologies and $\gamma$ values explored when full multipath (FMP) is combined with SIC either at $R$
or both at $R$ and $d$ it experiences by far higher average flow delay than all other flow allocation schemes discussed.
For topology 1 in Fig. \ref{fig:wireless_scenario} and $\gamma$=0.5 for example, the flow delay observed in the simulation
results for FMP-SIC(R) and FMP-SIC(R,d) is 2133.9 and 2111.3 slots respectively.
However this is expected since FMP assigns traffic on paths an interference-unaware manner experiencing a large number of
failed packets due to low received SINR. Secondly it does not adjust the flow assigned on a path based on the one that can
be serviced by its bottleneck link, resulting thus in unstable queues at the relay nodes.
For the case of topology 1 with $\gamma$=0.5 mentioned above the throughput ratio at $R$ for FMP-SIC(R) and FMP-SIC(R,d)
is $3.680$ and $3.658$ respectively.

\section{Conclusion}

In this work we explore the gain in terms of throughput that can be achieved at the network level by combining multipath utilization
and successive interference cancelation (SIC) for random access wireless mesh networks with multi-packet reception capabilities.
More precisely we explore different variants of a distributed flow allocation scheme aimed at maximizing average aggregate flow throughput (AAT)
where interference is either treated as noise or SIC is employed.
The flow allocation scheme discussed achieves up to $15.2\%$ higher AAT  when combined with SIC instead of treating interference as noise
for an SINR threshold ($\gamma$) value equal to $0.5$. For larger $\gamma$ values this improvement either becomes negligible or lower AAT
is achieved for the considered topologies. This is due to the fact that the increased interference caused by links whose success probability is significantly
improved with SIC is not compensated by the gain in terms of throughput. Moreover the improvement in terms of throughput by employing SIC instead
of treating interference as noise increases with the asymmetry among interfering links.

Future extensions of this work will include the study of deploying SIC jointly with TOFRA on larger topologies. However, one of the main challenges is
how to consider SIC in larger topologies when there are more than one interfering nodes to decode. Furthermore it would be of interest
to study the case where a receiver works on a hybrid manner by treating interference as noise or applying SIC based on the conditions. 

\section*{Acknowledgement}
M. Ploumidis was supported by HERACLEITUS II - University of Crete, NSRF (ESPA) (2007-2013) and is co-funded by the European Union and national resources.
The research leading to these results has also received funding from the People Programme (Marie Curie Actions) of the European Union's Seventh Framework Programme FP7/2007-2013/ under REA grant agreements n$^o$ [324515] (MESH-WISE) and n$^o$ [612361] (SOrBet).

\bibliographystyle{ieeetr}
\bibliography{IEEEabrv,bibliography}

\begin{thebibliography}{10}

\bibitem{6133896}
L.~Le, ``Multipath routing design for wireless mesh networks,'' in {\em Global
  Telecommunications Conference (GLOBECOM 2011), 2011 IEEE}, pp.~1 --6, dec.
  2011.

\bibitem{1717611}
M.~Alicherry, R.~Bhatia, and L.~E. Li, ``Joint channel assignment and routing
  for throughput optimization in multiradio wireless mesh networks,'' {\em
  Selected Areas in Communications, IEEE Journal on}, vol.~24, pp.~1960--1971,
  Nov 2006.

\bibitem{5501845}
G.~Middleton, B.~Aazhang, and J.~Lilleberg, ``Efficient resource allocation and
  interference management for streaming multiflow wireless networks,'' in {\em
  Communications (ICC), 2010 IEEE International Conference on}, pp.~1--5, May
  2010.

\bibitem{6130552}
G.~Middleton, B.~Aazhang, and J.~Lilleberg, ``A flexible framework for
  polynomial-time resource allocation in streaming multiflow wireless
  networks,'' {\em Wireless Communications, IEEE Transactions on}, vol.~11,
  pp.~952--963, March 2012.

\bibitem{1430253}
X.~Lin and N.~Shroff, ``Joint rate control and scheduling in multihop wireless
  networks,'' in {\em Decision and Control, 2004. CDC. 43rd IEEE Conference
  on}, vol.~2, pp.~1484--1489 Vol.2, Dec 2004.

\bibitem{4509706}
U.~Akyol, M.~Andrews, P.~Gupta, J.~Hobby, I.~Saniee, and A.~Stolyar, ``Joint
  scheduling and congestion control in mobile ad-hoc networks,'' in {\em
  INFOCOM 2008. The 27th Conference on Computer Communications. IEEE}, pp.~619
  --627, april 2008.

\bibitem{conf_icc_QiuBX12}
F.~Qiu, J.~Bai, and Y.~Xue, ``Towards optimal rate allocation in multi-hop
  wireless networks with delay constraints: A double-price approach.,'' in {\em
  ICC\/} \cite{conf_icc_QiuBX12}, pp.~5280--5285.

\bibitem{4712692}
P.~Wang, H.~Jiang, W.~Zhuang, and H.~Poor, ``Redefinition of max-min fairness
  in multi-hop wireless networks,'' {\em Wireless Communications, IEEE
  Transactions on}, vol.~7, pp.~4786 --4791, december 2008.

\bibitem{5089987}
Q.~Gao, J.~Zhang, and S.~Hanly, ``Cross-layer rate control in wireless networks
  with lossy links: leaky-pipe flow, effective network utility maximization and
  hop-by-hop algorithms,'' {\em Wireless Communications, IEEE Transactions on},
  vol.~8, pp.~3068 --3076, june 2009.

\bibitem{1603389}
J.-W. Lee, M.~Chiang, and A.~Calderbank, ``Jointly optimal congestion and
  contention control based on network utility maximization,'' {\em
  Communications Letters, IEEE}, vol.~10, pp.~216--218, Mar 2006.

\bibitem{Wang:2005:CRC:1062689.1062710}
X.~Wang and K.~Kar, ``Cross-layer rate control for end-to-end proportional
  fairness in wireless networks with random access,'' in {\em Proceedings of
  the 6th ACM International Symposium on Mobile Ad Hoc Networking and
  Computing}, MobiHoc '05, (New York, NY, USA), pp.~157--168, ACM, 2005.

\bibitem{7033238}
M.~Ploumidis, N.~Pappas, and A.~Traganitis, ``On the delay of a throughput
  optimal flow allocation scheme for random access wmns,'' in {\em Computer
  Aided Modeling and Design of Communication Links and Networks (CAMAD), 2014
  IEEE 19th International Workshop on}, pp.~218--223, Dec 2014.

\bibitem{b:tsirig1}
A.~Tsirigos and Z.~Haas, ``Analysis of multipath routing, part 1: The effect of
  packet delivery ratio,'' in {\em IEEE Transactions on Wireless Communications
  vol.3 no. 1}, pp.~138--146, 2004.

\bibitem{6134700}
P.-Y. Chen, W.-C. Ao, and K.-C. Chen, ``Rate-delay enhanced multipath
  transmission scheme via network coding in multihop networks,'' {\em
  Communications Letters, IEEE}, vol.~16, no.~3, pp.~281--283, 2012.

\bibitem{6335387}
M.~Ploumidis, N.~Pappas, V.~Siris, and A.~Traganitis, ``Evaluating forwarding
  schemes exploiting path diversity and degrees of redundancy in a realistic
  wireless environment,'' in {\em Computer Aided Modeling and Design of
  Communication Links and Networks (CAMAD), 2012 IEEE 17th International
  Workshop on}, pp.~95--99, 2012.

\bibitem{ref:papas_end_to_end}
N.~Pappas, V.~A. Siris, and A.~Traganitis, ``Path diversity gain with network
  coding and multipath transmission in wireless mesh networks,'' in {\em 2010
  IEEE International Symposium on a World of Wireless Mobile and Multimedia
  Networks (WoWMoM)}, pp.~1 --6, june 2010.

\bibitem{ref:papas_hop_by_hop}
N.~Pappas, V.~Siris, and A.~Traganitis, ``Delay and throughput of network
  coding with path redundancy for wireless mesh networks,'' in {\em Wireless
  and Mobile Networking Conference (WMNC), 2010 Third Joint IFIP}, pp.~1 --6,
  oct. 2010.

\bibitem{b:Pappas-arXiv-full-duplex}
N.~Pappas, M.~Kountouris, A.~Ephremides, and A.~Traganitis, ``Relay-assisted
  multiple access with full-duplex multi-packet reception,'' in {\em
  arXiv:1310.2773v4 [cs.IT]}, 2013.

\bibitem{b:Papadimitriou-arXiv}
G.~Papadimitriou, N.~Pappas, V.~Angelakis, and A.~Traganitis, ``Network-level
  performance evaluation of a two-relay cooperative random access wireless
  system,'' in {\em arXiv:1406.5949v2 [cs.NI]}, 2014.

\bibitem{4276942}
S.~P. Weber, J.~Andrews, X.~Yang, and G.~de~Veciana, ``Transmission capacity of
  wireless ad hoc networks with successive interference cancellation,'' {\em
  Information Theory, IEEE Transactions on}, vol.~53, pp.~2799--2814, Aug 2007.

\bibitem{Verdu:1998:MD:521411}
S.~Verdu, {\em Multiuser Detection}.
\newblock New York, NY, USA: Cambridge University Press, 1st~ed., 1998.

\bibitem{1210740}
S.~Toumpis and A.~Goldsmith, ``Capacity regions for wireless ad hoc networks,''
  {\em Wireless Communications, IEEE Transactions on}, vol.~2, pp.~736--748,
  July 2003.

\bibitem{5937210}
V.~Angelakis, L.~Chen, and D.~Yuan, ``Optimal and collaborative rate selection
  for interference cancellation in wireless networks,'' {\em Communications
  Letters, IEEE}, vol.~15, pp.~819--821, August 2011.

\bibitem{6339119}
D.~Yuan, V.~Angelakis, L.~Chen, E.~Karipidis, and E.~Larsson, ``On optimal link
  activation with interference cancelation in wireless networking,'' {\em
  Vehicular Technology, IEEE Transactions on}, vol.~62, pp.~939--945, Feb 2013.

\bibitem{5960833}
P.~Mitran, C.~Rosenberg, and S.~Shabdanov, ``Throughput optimization in
  wireless multihop networks with successive interference cancellation,'' in
  {\em Wireless Telecommunications Symposium (WTS), 2011}, pp.~1--7, April
  2011.

\bibitem{b:TITHaenggi}
X.~Zhang and M.~Haenggi, ``The performance of successive interference
  cancellation in random wireless networks,'' {\em IEEE Transactions on
  Information Theory}, vol.~60, pp.~6368--6388, Oct 2014.

\bibitem{b:SICSurvey}
N.~Miridakis and D.~Vergados, ``A survey on the successive interference
  cancellation performance for single-antenna and multiple-antenna ofdm
  systems,'' {\em IEEE Communications Surveys Tutorials}, vol.~15,
  pp.~312--335, First 2013.

\bibitem{b:TMC13}
S.~Sen, N.~Santhapuri, R.~Choudhury, and S.~Nelakuditi, ``Successive
  interference cancellation: Carving out mac layer opportunities,'' {\em IEEE
  Transactions on Mobile Computing}, vol.~12, pp.~346--357, Feb 2013.

\bibitem{b:PappasITW13}
N.~Pappas, M.~Kountouris, and A.~Ephremides, ``The stability region of the
  two-user interference channel,'' in {\em 2013 IEEE Information Theory
  Workshop (ITW)}, pp.~1--5, Sept 2013.

\bibitem{6824997}
M.~Ploumidis, N.~Pappas, and A.~Traganitis, ``Throughput optimal flow
  allocation on multiple paths for random access wireless multi-hop networks,''
  in {\em Globecom Workshops (GC Wkshps), 2013 IEEE}, pp.~263--268, Dec 2013.

\bibitem{DBLP:journals/corr/PloumidisPT14}
M.~Ploumidis, N.~Pappas, and A.~Traganitis, ``Tofra: Throughput optimal flow
  rate allocation with bounded delay for random access wireless mesh
  networks,'' {\em CoRR}, vol.~abs/1406.6304, 2014.

\bibitem{Pappas:2014:SPI:2611842.2612059}
N.~Pappas, A.~Ephremides, and A.~Traganitis, ``Stability and performance issues
  of a relay assisted multiple access scheme with mpr capabilities,'' {\em
  Comput. Commun.}, vol.~42, pp.~70--76, Apr. 2014.

\bibitem{b:Tse}
D.~Tse and P.~Viswanath, {\em Fundamentals of wireless communication}.
\newblock New York, NY, USA: Cambridge University Press, 2005.

\bibitem{ref:ns2}
``{The Network Simulator NS-2}.'' \url{http://www.isi.edu/nsnam/ns/}.

\bibitem{ref:dei80211mr}
S.~research group, ``dei80211mr: a new 802.11 implementation for ns-2,'' 2008.
\newblock Available at http://www.dei.unipd.it/wdyn/?IDsezione=5090. [Último
  acesso: 18/01/2010].

\end{thebibliography}

\end{document}